\documentclass[pre,aps,amsmath,amssymb,superscriptaddress,notitlepage,twocolumn]{revtex4-2}
\usepackage[colorlinks,bookmarks=false,citecolor=blue,linkcolor=blue,urlcolor=blue]{hyperref}
\usepackage[all]{hypcap}   

\usepackage{amsmath,amssymb}
\usepackage{graphicx}
\graphicspath{{figures/}{/}}
\usepackage{dcolumn}
\usepackage{bm}
\usepackage[capitalise,compress]{cleveref}
\crefname{section}{Sec.}{Secs.}
\Crefname{section}{Section}{Sections}

\usepackage{verbatim}
\usepackage{color}
\usepackage{soul}
\usepackage{placeins}  
\usepackage{flafter}     
\usepackage{color}
\usepackage{hhline}
\usepackage{physics}
\usepackage{hyperref}

\usepackage{float}
\usepackage{caption}
\usepackage{subcaption}
\usepackage{lipsum} 

\usepackage[mathscr]{eucal}
\usepackage{tikz}
\usepackage{xcolor}

\newcommand{\bo}{\begin{outline}}
\newcommand{\eo}{\end{outline}}

\newcommand{\qed}{\nobreak \ifvmode \relax \else
      \ifdim\lastskip<1.5em \hskip-\lastskip
      \hskip1.5em plus0em minus0.5em \fi \nobreak
      \vrule height0.75em width0.5em depth0.25em\fi}
\begin{document} 

\title{Dichotomy in the effect of chaos on ergotropy}

\author{Sreeram PG}
\email{sreerampg7@gmail.com}
\affiliation{Department of Physics, Indian Institute of Science Education and Research, Pune 411008, India}
\author{J. Bharathi Kannan}
\email{bharathikannan1130@gmail.com}
\affiliation{Department of Physics, Indian Institute of Science Education and Research, Pune 411008, India}
\author{S. Harshini Tekur}
\affiliation{Department of Physics, Indian Institute of Science Education and Research, Pune 411008, India}
\author{M. S. Santhanam}
\email{santh@iiserpune.ac.in}
\affiliation{Department of Physics, Indian Institute of Science Education and Research, Pune 411008, India}


\begin{abstract}
{The maximum unitarily extractable work from a quantum system -- ergotropy -- is the basic principle behind quantum batteries, a rapidly emerging field. This work studies ergotropy in two quantum chaotic systems, the quantum kicked top and the kicked Ising spin chain, to illustrate the effects of chaotic dynamics. In an ancilla-assisted scenario, chaos enhances ergotropy when the state is known, a consequence of large entanglement production in the chaotic regime. When the state is unknown, we need to at least partially characterize the state using coarse-grained measurements for useful extraction of work. In this case,  chaos impedes ergotropy by suppressing information gained from coarse-grained measurements, while entanglement with an ancilla still facilitates ergotropy. In this scenario, we study the interplay between chaos and entanglement and find a sweet spot in the chaos parameter for optimal work. Our results point to the potential of quantum chaos-assisted batteries for better work extraction.} 
\end{abstract}

\maketitle
In the last two centuries, thermodynamics has established the relation between work and energy in classical systems \cite{Greiner}.  
How thermodynamics emerges in finite-size quantum systems and its relation to quantum correlations is currently the central theme of quantum thermodynamics \cite{BinCorGog2018,goold2016role,vinjanampathy2016quantum}. For classical systems in contact with thermal bath, maximal work extractable at constant temperature is bounded by the change in free energy. For an isolated and finite quantum system, {\it ergotropy} is the analogous quantity and represents optimal work extractable through unitary operations from a quantum state \cite{allahverdyan2004maximal}. 

Naturally, ergotropy has been widely applied in the context of quantum batteries whose main operations -- charging and discharging -- are related to storing energy and releasing it for doing work \cite{CamGheQua2024,binder2015quantacell,alicki2013entanglement,campaioli2017enhancing,monsel2020energetic,ZhaWanYua2024}.
In tune with current interest in quantum information theory, a question of fundamental interest is how ergotropy is affected by quantum coherence \cite{francica2020quantum,ShiDinWan2022,MaXuLi2024} and quantum correlations such as entanglement \cite{alicki2013entanglement,alimuddin2019bound,JosAliMah2024,Francica2022,TouBarDef2021,mitra2024sunburst}, in systems with finite size baths \cite{Lobejko2021}. In general, quantum correlations have been shown to enhance ergotropy \cite{alicki2013entanglement,francica2017daemonic,francica2020quantum,alimuddin2019bound}. This was experimentally observed in a quantum device with entangled ions as a working medium \cite{ZhaWanYua2024}, and in a single electron spin of a nitrogen-vacancy center \cite{NiuWuWan2024}. Further, quantum correlations also enhance the charging performance of batteries \cite{MaXuLi2024}. Although optimal work may also be extractable without the aid of quantum correlations, usually it requires more operations \cite{hovhannisyan2013entanglement}. 


One approach to studying the effect of entanglement production on ergotropy is to look at quantum chaotic systems. In such systems, more classical
chaos generation is usually associated with more entanglement generation until it saturates the limit of nearly
complete chaos in the system. Thus, one key feature of chaotic quantum systems is the production of large quantum correlations \cite{wang2004entanglement,madhok2015signatures,madhok2018quantum}. Based on this discussion, one expects chaos to aid and enhance ergotropy. 

On the other hand, the effectiveness of quantum machines is limited by error propagation and loss of quantum control, with presence of chaos aggravating these issues \cite{georgeot2000emergence,preskill2018quantum, arute2019quantum,berke2022transmon,basilewitsch2023chaotic}. From a classical viewpoint, the free energy available to convert into thermodynamic work is lower when the system has a large entropy due to chaos. A quantum thermodynamic process requiring state characterization is one occasion where chaos could be disadvantageous for ergotropy because of the entropy production.
When the initial state is unknown, its characterization by tomography is required before one can extract work from it \cite{vsafranek2023work}. This leads to an interesting dichotomy about the role of chaos vis-a-vis ergotropy. It is now difficult to decide {\it apriori} whether chaos is useful for ergotropy or not. In this paper, using quantum chaotic models, we show that ergotropy benefits from chaos when the state is known. However, when the state is unknown and requires reconstruction, then chaos is not helpful. In the latter case, a competition between these two faces of chaos -- aiding entanglement and inhibiting state reconstruction -- results in optimal ergotropy that is strongly correlated with coarse-grained entropy measures.

Given a quantum state $\rho$ and a reference Hamiltonian $H$, ergotropy is the maximal work extractable through unitary transformation and can be written as \cite{allahverdyan2004maximal,alicki2013entanglement,CamGheQua2024}
\begin{equation}
    W(\rho,H)= \Tr(\rho H)-\mathrm{min}_\mathcal{U}\Tr(\mathcal{U} \rho \mathcal{U}^\dagger H), 
    \label{Eq: ergotropy}
\end{equation}
where the minimum is taken over all unitary transformations $\mathcal{U}$. 
The unique state $\pi= \mathcal{U} \rho \mathcal{U}^\dagger$ with respect to $(\rho, H)$ which satisfies Eq. (\ref{Eq: ergotropy})
is called the passive state since no work can be extracted from $\pi$ through any cyclical variation of a parameter of the Hamiltonian. This motivates a working formula for ergotropy in terms of eigen decompositions: $H= \sum_k \epsilon_k \ket{\epsilon_k}\bra{\epsilon_k}$, where $\epsilon_k \leq \epsilon_{k+1}$ and $\rho= \sum_k r_k \ket{r_k}\bra{r_k},$ where $r_k \geq r_{k+1}$. Then, we have
$\pi= \sum_k r_k \ket{\epsilon_k}\bra{\epsilon_k}$~\cite{alicki2013entanglement}, and using these, Eq. (\ref{Eq: ergotropy}) can now be rewritten as \cite{francica2020quantum}
\begin{equation}
    W(\rho,H)= \sum_k \epsilon_k(\rho_{kk}-r_k), 
    \label{Eq: work}
\end{equation}
where $\rho_{kk}=\sum_{k'} r_{k'} |\bra{r_{k'}}\ket{\epsilon_k}|^2$ is the fraction of $\rho$ in the energy eigenstate $\ket{\epsilon_k}.$
Physically, the work extraction process brings the initial state to a lower energy state with respect to $H$ and garners the difference in energy.  No energy can be extracted if the initial state is $\pi$. To elucidate the role of chaos, we study ergotropy using two quantum chaotic systems, namely, the quantum kicked top and a kicked Ising spin chain.

\section{Quantum kicked top and the kicked Ising model}
The quantum kicked top is a well-studied model of quantum chaos, whose dynamics is characterized by the spin-angular momentum vector ${J}=(J_x,J_y,J_z)$~\cite{haake1987classical,haake1991quantum,chaudhury2009quantum}.
As a time-periodic system, its quantum dynamics is described by the period-1 Floquet operator ($\hbar=1$)
$
U = \exp (-i \frac{\kappa}{2j} J_z^2 ) \exp(-i \alpha  J_y)$,
where $\kappa$ is the kick strength,  $\alpha=\pi/2$ is the angle of precession about $y$-axis and $j$ the spin angular momentum. {The kicking parameter $\kappa$ determines the strength of the nonlinear rotation term $e^{\frac{-i\kappa j_z^2}{2j}}$, which induces chaotic behaviour by disrupting the continuous linear precession about the orthogonal axis. This creates complex dynamics when the kick strength is large, leading to chaos \cite{haake1987classical}.} The kicked top can also be viewed as an interacting multi-qubit system \cite{wang2004entanglement}, in which the spin-angular momentum is a collective operator composed of smaller spins. 

We consider a system-ancilla (denoted by subscripts $S$ and $A$) set-up with an initial state $\ket{\Psi(0)}= \ket{\psi_S} \otimes \ket{\psi_A},$ {where the subsystem states $\ket{\psi_S}$ and $\ket{\psi_A}$ are chosen to be Haar random, } and 
known to the experimenter. Then the system undergoes Floquet evolution for $t$ time steps: $\ket{\Psi(t)}= U^t\ket{\Psi(0)},$ where
\begin{equation}
    U=  \exp (-i \frac{\kappa}{2j} (J_{Sz}+J_{Az})^2 ) \exp(-i \alpha  (J_{Sy}+J_{Ay})). 
    \label{Eq. SA floquet}
\end{equation}
Here $J_{S\gamma}$ and $J_{A\gamma}$ denote the $\gamma$ component ($\gamma=y,z)$ of the spin operator, and $j=j_S+j_A$.
The resulting state $\ket{\Psi(t)}$ is no longer a product state since $U$ establishes quantum correlations between the system and ancilla. After the evolution, the system state $\rho_S$ can be obtained by partial-tracing the ancilla.

The kicked Ising spin chain \cite{prosen2000exact,prosen2002general, lakshminarayan2005multipartite,mishra2014resonance,pineda2007universal,herrmann2023characterizing} with open boundaries is governed by the period-1 Floquet operator
 \begin{equation}
     U= e^{-iH_{\mathrm{free}}/2}e^{-iH_{\mathrm{kick}}} e^{-iH_{\mathrm{free}}/2}. 
     \label{eq: ising}
 \end{equation}
 Here $H_{\mathrm{free}}= C \sum_{i=1}^{L-1} \sigma_{iz} \sigma_{(i+1)z},$ where $C$ is the coupling strength, and $L$ is the length of the spin chain.  In this, $H_{\mathrm{kick}}= M \sum_{i=1}^{L} (\cos{\Theta}_i\sigma_{ix}+\sin{\Theta_i}\sigma_{iz}),$ where $M$ is the strength of the magnetic field (kick term) which is turned on and off periodically. The tilt $\Theta_i$ for the $i^{th}$ spin determines the angle at which the magnetic field acts on it in the $x$-$z$ plane. The length of the spin-chain is set to $L=8$. In the system-ancilla picture, the system consists of $6$ spins and the remaining two spins to comprise the ancilla. The other parameters are $C=0.8$ and, following  \cite{herrmann2023characterizing}, $\{\Theta_i \}=\{7,7,8,8,8,8,7,7\}\pi/32$.

\section{Ergotropy of known states with ancilla measurements}: First, we consider the case of  known initial state $\rho_S$. Then, the extractable work is $W(\rho_S,H_S)$ (see Eq. \ref{Eq: work}), where a unitary transformation takes the system to the passive state. Further, $W(\rho_S,H_S)$ can be enhanced by coupling the system to an ancilla and then performing ancilla measurements \cite{francica2017daemonic}. Let $\{\Pi^a_A\}$ be the complete set of mutually orthogonal projectors, where $A$ denotes the ancilla and $a$ the outcome. After ancilla is measured and given the outcome $a$, the post-measurement state of the system is given by 
\begin{equation}
    \rho_{S|a}= \Tr\left(\Pi^a_A\ketbra{\Psi(t)} \Pi^a_A\right)/p^a,
\end{equation}
where $p^a= \Tr(\Pi^a_A \ketbra{\Psi(t)})$ is the probability for outcome $a$. As the system and the ancilla are entangled, each measurement outcome gives us more information about the system state.
The extractable work is given by \cite{francica2017daemonic}
\begin{multline}
    W_{\{\Pi_A\}}(\rho_S,H_S)= \Tr(\rho_SH_S) -\\ \sum_a p^a \mathrm{min}_{\mathcal{U}_S}\Tr(\mathcal{U}_S\rho_{S|a}\mathcal{U}_S^\dagger H_S) .\label{Eq: daemonic}
\end{multline}
In this, the second term represents an average over different outcomes for the energy content of the passive state. 

\begin{figure}[t]
    \centering
    \includegraphics[width=\linewidth]{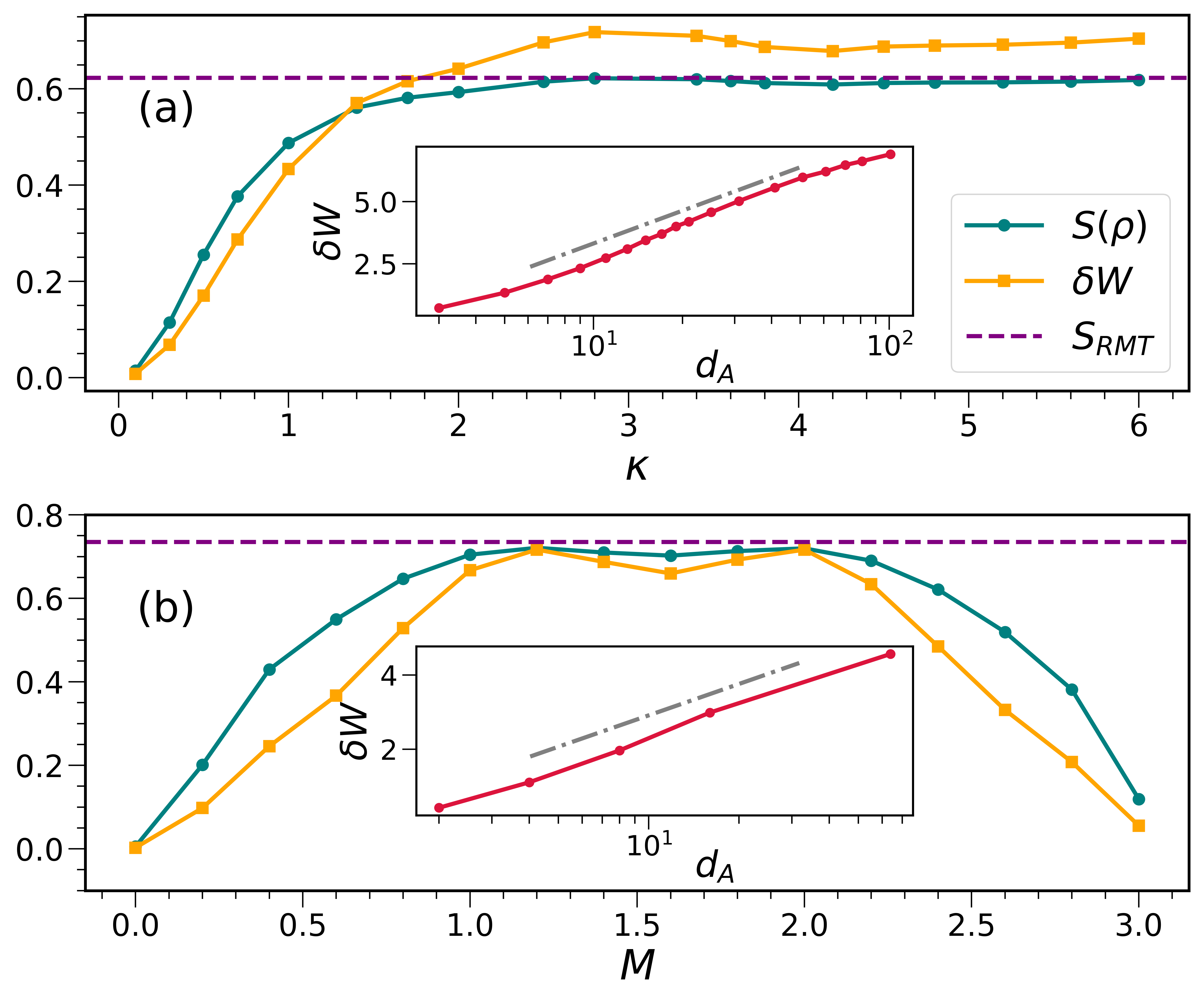}
    \caption{Entanglement entropy $S(\rho)$ and work gain $\delta W$ with ancilla measurements as a function of the kick strength for (a) the kicked top and (b) the kicked Ising model.  $\delta W$ closely follows $S(\rho)$. In the chaotic regime,  $S(\rho)$ converges to the random matrix average $S_\mathrm{RMT}$ (dashed lines). For the kicked top, $S_\mathrm{RMT} \approx 0.6229$, and for the kicked Ising chain, $S_\mathrm{RMT} \approx 0.7354$.  $\delta W$ in the chaotic regime exhibits logarithmic growth with $d_A$ (Inset).}
    \label{Fig:ent_work}
\end{figure}

\subsection{Demonstration of the ergotropy of known states}
 The central quantity of interest is the work gain  $\delta W= W_{\{\Pi_A\}}(\rho_S,H_S)- W(\rho_S,H_S)$ and its variation as a function of kick strength. For both kicked top and Ising dynamics, in the regime of chaotic dynamics, large entanglement is generated \cite{haake1987classical,wang2004entanglement,kumari2019untangling}. Without loss of generality, the system of interest is fixed to be $H_S=-J_{Sz}$.{An ensemble of $10^3$ initial product states of the system and ancilla,  each chosen uniformly at random are evolved for $t=3$ time steps.}  Although the trends in Fig. \ref{Fig:ent_work} are independent of the choice of time,  we choose $t=3$ in the scrambling regime for better resolution on the amount of chaos.  

For kicked top, we choose  $j_S=19/2$ (system dimension $d_S=20$),  and $j_A=1$ (ancilla dimension $d_A=3$). It is integrable at $\kappa=0$, and progressively becomes chaotic as $\kappa \gg 1$. {{For the kicked Ising chain, the system consists of 6 spins, and the corresponding $H_S$ is the $J_Z$ operator in $2^6$ dimensions.}} Chaotic dynamics, at early times, is usually associated entanglement generation, while it saturates at late times for finite systems. As Fig. \ref{Fig:ent_work}(a) shows, $\delta W$ closely follows the entanglement evolution with $\kappa$. In Fig. \ref{Fig:ent_work}(b), similar results are observed for the kicked Ising model as well. The Kicked Ising model is integrable at $M=0$ and the entanglement remains small.  As $M$ increases, more chaos leads to more entanglement generation, as shown in Fig. \ref{Fig:ent_work}(b). The eigenvalue spacing distribution changes from Poisson to Wigner-Dyson statistics in this regime 
\cite{supp}. 
Unlike the kicked top, this system shows the revival of the near-integrable regime as $M \to \pi$.  This feature is reflected in both the entanglement between the subsystems and  work gain $\delta W$. In both the systems,  maximal entanglement 
numerically equals the random matrix average  $S_{\rm RMT}=1- \frac{d_S+d_A}{1 + d_S d_A}$ \cite{lubkin1978entropy} (dashed line in Fig. \ref{Fig:ent_work}). 
Further, as seen in the inset of Fig. \ref{Fig:ent_work}, for a fixed value of $t=3$ and system size,  we observe that $\delta W \propto \log d_A$ in the chaotic regime. This behaviour is reminiscent of the growth of entanglement with subsystem size \cite{page1993average}.

\subsection{The case of three subsystems and usable entanglement} 
\begin{figure}
    \centering
    \includegraphics[width=\linewidth]{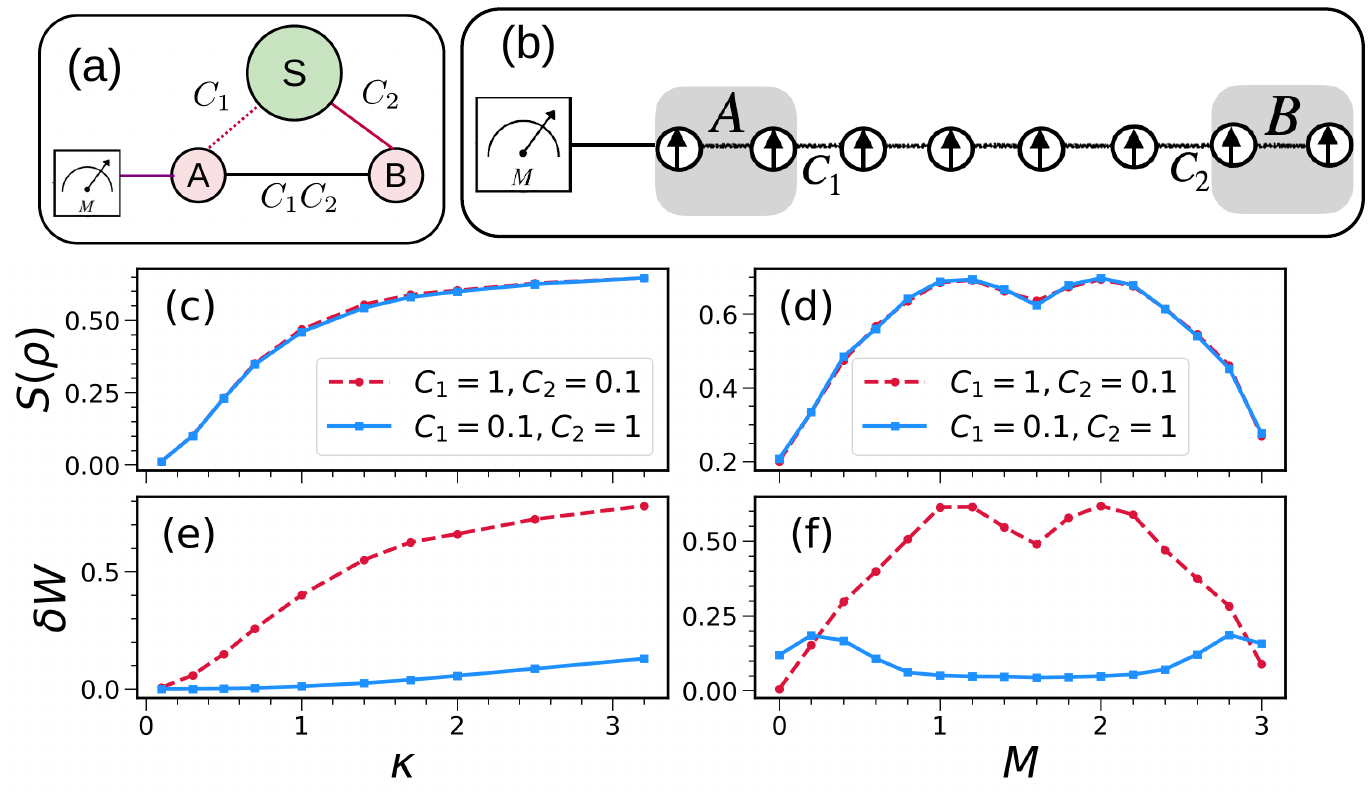}
    \caption{Schematic of the interaction between the system ($S$), ancilla ($A$), and auxiliary ($B$) subsystems for (a) the kicked top and (b) the kicked Ising model.
    (c) and (d) show the entanglement entropy of the system $S$ as a function of the kick strength.
    (e) and (f) show that   $\delta W$ differs significantly based on how the entanglement is distributed.}
    \label{fig:3 subsystem}
\end{figure}

How does entanglement sharing affect the work gain in a general case with more than two subsystems? We answer this question by considering a schematic shown in Fig. \ref{fig:3 subsystem}(a). In this scheme, the system $S$ is coupled to two subsystems: an ancilla $A$ which is measured, and an unmeasured auxiliary system $B$ with coupling strengths $C_1$ and $C_2$, respectively.
The correlations between the subsystems can be tuned by varying $\{C_1, C_2\}$ so that the exponent in the torsion term for the kicked top is modified to  $(J_{Sz}+ C_1\:J_{Az}+C_2\:J_{Bz})^2$.
Similarly, for the kicked Ising chain, we introduce variable coupling denoted by $C_1$ and $C_2$ at the edge states identified as ancilla $A,$ and the auxiliary system $B$ respectively, as shown in the schematic  Fig. \ref{fig:3 subsystem}(b).  

With this model, we have a system whose entanglement is the same for different choices of $C_1$ and $C_2$ (Fig. \ref{fig:3 subsystem}(c,d)). Yet, the ergotropy in these two configurations can be different as seen in Fig. \ref{fig:3 subsystem}(e,f). This result implies that only the entanglement with the measured subsystems affects ergotropy. Generalization of the three-subsystem ergotropy to more subsystems is straightforward. 

\section{Ergotropy of unknown states}
What if the state $\rho_S$ from which work needs to be extracted is not known? 
As before, a reduced state $\rho_{S|a}$ is obtained post the ancilla measurement,. However, the experimenter has to perform measurements on $\rho_{S|a}$ to determine the state before transforming it to the corresponding passive state. This additional step of full tomography is expensive, particularly for large system dimensions. Hence, we limit our approach to coarse-grained measurements of the system.

Consider a $d$-dimensional Hilbert space $\mathcal{H}^S$ of a quantum system, partitioned into orthogonal subspaces (macrostates)  $\lbrace \mathcal{H}^S_i\rbrace,$ such that $\mathcal{H}^S= \oplus_i^d \mathcal{H}^S_i$. Let $\Pi_i$ denote the projection onto a macrostate $\mathcal{H}^S_i$. 
In practice, $\Pi_i$ can be constructed by combining eigenvectors of an observable that the experimenter can measure. If projections are the only measurements that can be performed, then the set of projections $\chi = \lbrace \Pi_i \rbrace$ represents a \textit{coarse-graining}. The dimension of a macrostate is the coarse-graining length $V_i=\Tr \Pi_i$. For uniform coarse-graining, $V_i=n \: \forall i, \: n\in\mathbb{Z}^+,$ and is denoted by $\chi_n$. We use two different measurement protocols to extract ergotopy, as discussed below.

\textit{Protocol-1}: Suppose that the collective system-ancilla state is denoted by $\rho.$ First, a projective measurement on the ancilla is performed. For an outcome $a$ of the ancilla, let us denote the corresponding reduced state of the system as $\rho_{S|a}.$ Since we do not know this state, there is no way to know the corresponding passive state. Therefore, to extract energy from the system, we do coarse-grained measurements to gain partial information about the system.  

Upon coarse-grained projective measurement, the probability to find $\rho_{S|a}$ in a macrostate $\mathcal{H}^{S}_i$ is $p_i=\Tr(\Pi_i \rho_{S|a})$, and the reconstructed state  using the coarse-measurement information is 
\begin{equation}
     {\rho^{rc}_{S|a}}= \sum_i p_i \Pi_i/V_i.
\end{equation}
If $V_i>1,$ the measurements are coarse-grained and the fidelity $F(\rho^{rc}_{S|a},\rho_{S|a})<1.$ The remaining amount of uncertainty  about the system state after the coarse measurements is quantified by \textit{observational entropy} (OE), defined as \cite{vsafranek2019quantum, vsafranek2019quantum1, vsafranek2021brief, strasberg2021first}
\begin{equation}
\mathbb{H}_\chi (\rho_{S|a}) = -\sum_i p_i \log \frac{p_i }{V_i}. 
\label{eq:oe} 
\end{equation}
Now, to find the ergotropy from $\rho_{S|a}$, we can evolve the system to its unitarily achievable unique passive state denoted by $\pi^{rc}_{S|a}.$  Repeating this procedure,  we can find the unique passive state of the reduced system state corresponding to each outcome of the ancilla.
The work gained using this protocol is
\begin{equation}
    W^{rc}_{\{\Pi_A\}}(\rho_S,H_S)= \Tr(\rho_SH_S) - \sum_a p^a \Tr(\pi^{rc}_{S|a} H_S),
    \label{Eq: daemonic_ent}
\end{equation}
where $p^a$ denotes the probability for the ancilla outcome $a.$

\textit{Protocol-2}: In this protocol, we change the order of averaging. First, we average over the ancilla measurements and obtain a statistically averaged reduced state 
\begin{equation}
    \rho^{rc}_S= \sum_a p^a \rho^{rc}_{S|a}.
\end{equation} 
Since this is an effective reduced state that lacks correlations to individual ancillary outcomes, the information content about the system is lower (as compared to protocol-1). The unknown state $\rho^{rc}_S$ is then reconstructed with coarse measurements and taken to its passive state $\pi^{rc}_S$. The ergotropy is now given by
\begin{equation}
      \overline{W}^{rc}_{\{\Pi_A\}}= \Tr(\rho_SH_S) -  \Tr(\pi^{rc}_{S} H_S). 
      \label{daemonic 2}
\end{equation}
 Note that the passive states obtained with coarse-grained measurements in the above protocols are farther from the actual passive state, leading to lower ergotropy \cite{vsafranek2023work,vsafranek2023ergotropic}. 
Further, in protocol-2, averaging process destroys entanglement. Therefore, only the detrimental effect of chaos survives, and the passive state obtained is even poorer in quality.

\begin{figure}
\includegraphics[width=1\linewidth]{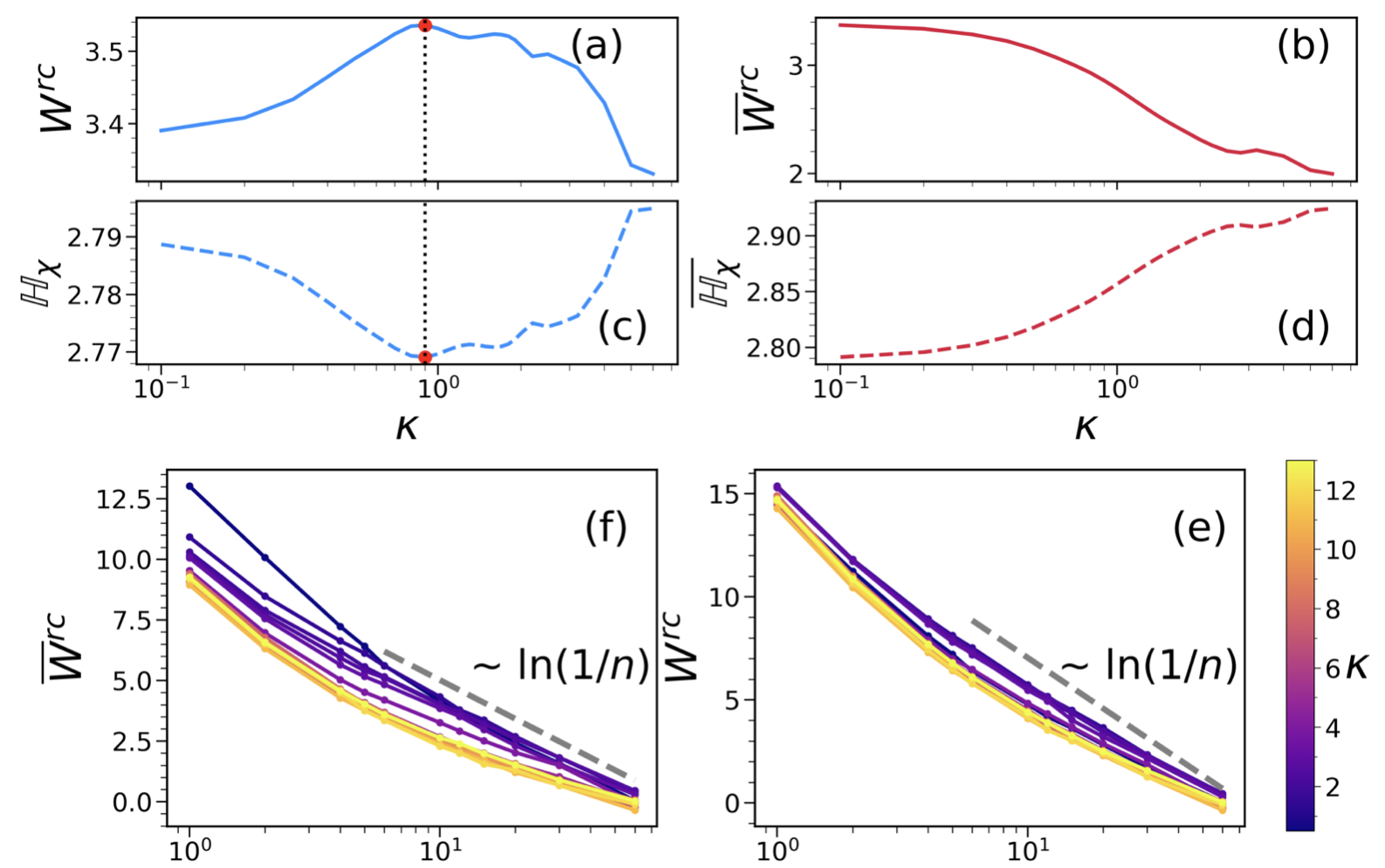}
     \caption{Unknown states of kicked top system. (a) ergotropy and (c) OE plotted against $\kappa$ using protocol-1 (ancilla outcomes used to gain information about the system). (b) ergotropy, and (d) OE using protocol-2 (average state used after ancilla measurements). The coarse-graining is fixed at $\chi_2$.   
     (e) Ergotropy decreases with coarse-graining length $n$. When entanglement plays a role, the decay is slower than $\log (1/n)$ at large $n$.  (f) Without entanglement, ergotropy decays as $\log (1/n)$ for small $\kappa$, consistent with the growth of ignorance \cite{pg2023witnessing}.}
     \label{fig: coarse_work}
\end{figure}

\subsection{Demonstration of the ergotropy of unknown states}
Firstly, let us consider protocol$-1$. Figure \ref{fig: coarse_work}(a) displays ergotropy obtained from $\rho^{rc}_{S}$ as a function of $\kappa$ for the kicked top. Two competing effects -- entanglement generation and state reconstruction -- come into play. As evident in Fig. \ref{fig: coarse_work}(a), $W^{rc}$ increases until about $\kappa=1$ before it begins to decay. This can be understood as follows : when the state is unknown, ergotropy depends on the fidelity of state reconstructed through coarse measurements. Let $\pi$ and $\widetilde{\pi}$ represent the true (unknown) state and estimated state through coarse measurements, respectively.  Then, the fidelity of the reconstructed state can be denoted by $F(\pi,\widetilde{\pi})$. The information any observer gains from coarse measurements depends on  $F(\pi,\widetilde{\pi})$.  The observational entropy $\mathbb{H}_\chi$ (Eq. \ref{eq:oe}) is a suitable measure of uncertainty of a state subjected to coarse measurements. This argument posits that smaller $F(\pi,\widetilde{\pi})$ implies larger $\mathbb{H}_\chi$, and consequently smaller ergotropy. Based on strong numerical evidence in Fig. \ref{fig: coarse_work}(a,c), it is observed that $W^{rc} = b_1 - b_2 \mathbb{H}_\chi$, where $b_1$ and $b_2$ are dimensionful constants. Clearly, $\kappa$ at which maximum ergotropy is achieved corresponds to a minimum in $\mathbb{H}_\chi $  (Fig. \ref{fig: coarse_work}(c)). This result helps to compute maximal ergotropy using observational entropy for chaotic systems.

\begin{figure}[t]
    \centering
    \includegraphics[width=0.5 \textwidth]{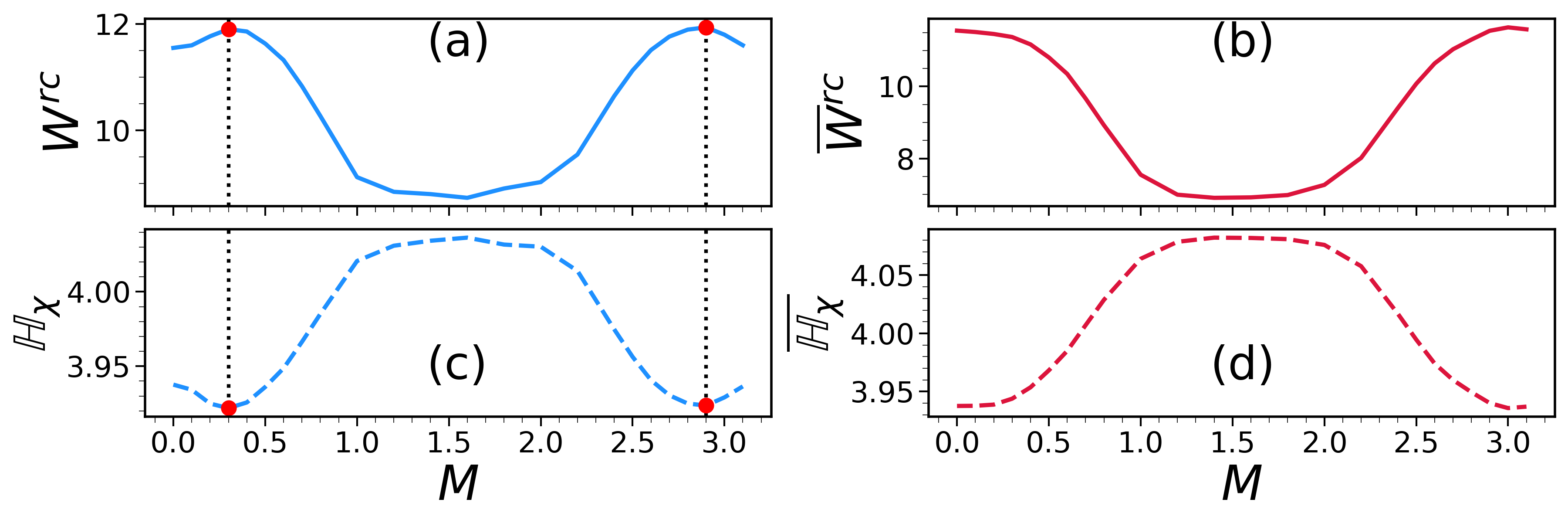}
    \caption{Kicked Ising model for unknown state and coarse-graining $\chi_2$. (a,b) Ergotropy, and (c,d) OE as a function of kick strength $M$. The left panel uses protocol-1, while the right panel is for protocol-2.} 
    \label{fig: coarse_ising}
\end{figure}

As observed in Fig. \ref{fig: coarse_work}(a), 
entanglement aids ergotropy until $\kappa \simeq 1$, and the effect of chaos is subdued (OE is high as seen in Fig. \ref{fig: coarse_work}(c)). Indeed, the classical kicked top also is in the near-integrable regime for $0 \le \kappa \le 1$. Aided by entanglement generation and lack of chaos leads to increasing ergotropy. As the system becomes more chaotic beyond $\kappa \ge 1$, coarse-grained measurements become less useful in state reconstruction and  $F(\pi,\tilde{\pi})$ drops. Therefore, the extractable work decreases. Effectively, under the opposing effects of entanglement and coarse-graining, ergotropy is non-monotonic, and an extrema exists. 

If protocol-2 is applied, entanglement is washed out in averaging. Then, the growth phase observed in Fig. \ref{fig: coarse_work}(a) should be absent. As expected, in the absence of entanglement, Fig. \ref{fig: coarse_work}(b) shows monotonic decrease of ergotropy, and this matches with the monotonic increase of $\mathbb{\bar H}_\chi$ seen in Fig. \ref{fig: coarse_work}(d). Thus, protocol-2 allows us to infer that entanglement and coarse graining are jointly responsible for the existence of maxima in ergotropy.
It might be emphasised that ergotropy (estimated through both protocols) depends on coarse graining length $n$. The ignorance about a state grows with $n$ as quantified by observational entropy and is $\mathbb{H}_\chi\propto \log 1/n$ (see Fig. 2 in Ref. \cite{pg2023witnessing}). Due to the effect of entanglement in Fig. \ref{fig: coarse_work}(e), the decay of $W^{rc}$ is slightly slower than $\log 1/n$. However, when entanglement is washed out, as seen in Fig \ref{fig: coarse_work}(f), for small $\kappa$ and at larger coarse-grainings, $\overline{W}^{rc} \propto \log 1/n$ consistent with Ref. \cite{pg2023witnessing}.

Figure \ref{fig: coarse_ising} shows ergotropy for the kicked Ising model estimated using Eq. (\ref{Eq: daemonic_ent}) and Eq. (\ref{daemonic 2}) through protocol-1 and -2. In Fig. \ref{fig: coarse_ising}(a,b), work is averaged over $10^{3}$  initial product states of Haar random subsystem states. As expected, $W^{rc} > \overline{W}^{rc}$, i.e, work extraction is more when entanglement plays a role than if it were suppressed through averaging over ancilla (protocol-2). The initial increase in $W^{rc}$ aided by entanglement, though small, can be noticed in Fig. \ref{fig: coarse_ising}(a). As $M$ increases, state reconstruction becomes difficult due to chaos, and work gain decreases. Once again, observational entropy predicts the position of extrema of ergotropy as a function of $M$, as shown in Fig. \ref{fig: coarse_ising}(c,d). The difference  $W^{rc} - \overline{W}^{rc}$ at any $M$ or $\kappa$ can be entirely attributed to the useful role of entanglement.


\section{Conclusion}
{
How does the presence of chaos affect ergotropy, or the maximal work extraction in a quantum system ? This question carries intrinsic interest as problem of quantum statistical physics, and is of practical interest
in the context of quantum batteries. This work illuminates dichotomy in the effects arising due to chaos. As chaos typically enhances entanglement, one might apriori expect enhancement in ergotropy as well. However, contrary to such expection, chaos plays a dual role. It is shown that chaos can assist or inhibit ergotropy, depending on the observer's knowledge of the system. If the observer has full knowledge of the state, and in an ancilla assisted situation, chaos tends to enhance ergotropy. When the state is not completely known, two competing effects come into play -- chaos aids ergotropy, but chaos also inhibits state characterization through coarse-grained measurements. This competition leads to the emergence of optimality, a sweet spot in the chaos parameter at which maximal work can be extracted. This has been demonstrated using two models; quantum kicked top and kicked Ising chain. Our preliminary investigations indicate that these results can be generalized using random matrix ensembles, and it will be taken up elsewhere. As a practical outcome, our results could lead to better batteries assisted by quantum chaos. In future, chaos-assisted battery needs further study, and the chaotic regime holding promise for the next generation of energy storage devices is of immense importance, especially since this regime is not considered to be conducive for quantum computing purposes.} 


\bibliography{ref}
\bibliographystyle{elsarticle-num}

\end{document}